\documentclass[cits]{PoS}
\usepackage{url}
\usepackage{amssymb,amsmath,amstext,amsthm,amsfonts}
\usepackage{graphicx}
\newcommand{\plotscale}{.60}
\newcommand{\carbon}{{\atom{12}{C}}}
\newcommand{\atom}[2]{\mbox{$^{#1}\text{#2}$}}
\newcommand{\oxygen}{{\atom{16}{O}}}

\newcommand{\refcite}[1]{Ref.~\cite{#1}}
\newcommand{\onepion}{one pion~}

\newcommand{\GeV}{\; \mathrm{GeV}}
\newcommand{\MeV}{\; \mathrm{MeV}}

\title{Neutrino-Long-Baseline Experiments\\ and Nuclear Physics}

\ShortTitle{Neutrino Interactions with Nuclei}

\author{\speaker{Ulrich Mosel} and Olga Lalakulich\thanks{Work supported by DFG}\\
        Institut fuer Theoretische Physik, Universitaet Giessen, D-35392 Giessen, Germany\\
        E-mail: \email{mosel@physik.uni-giessen.de}}

\abstract{Neutrino long-baseline experiments nowadays all use nuclear targets. The extraction of neutrino oscillation parameters from such experiments requires a good understanding of the interaction of neutrinos with nuclei. In this talk we discuss results on quasielastic scattering and pion production which are the relevant processes in the neutrino energy regime around 1 GeV. We also discuss implications of the reaction mechanisms for the reconstruction of the neutrino energy which is not known a priori.}

\FullConference{50th International Winter Meeting on Nuclear Physics - Bormio2012,\\
		23-27 January 2012\\
		Bormio, Italy}

\begin{document}

\section{Introduction}

This is the write-up of a talk that one of the authors (U.M.) gave on the occasion of the 50th Winter Workshop on Nuclear Dynamics in Bormio\footnote{The author dedicates this talk to Ileana Iori who played the central role in bringing this Bormio meeting together year after year. She also supported many generations of my graduate students and made it possible for them to experience the unique atmosphere at Bormio. In her memory this talk was given.}. Since this meeting is attended mainly by scientists working in very different fields, mostly high-energy heavy-ion physics and hadron physics, this talk tries to give an impression of what nuclear physics can contribute to neutrino physics, in particular with long baseline (LBL) beams. This article thus tries to review some of our work in this field for non-experts; as such it draws heavily on our former publications \cite{Leitner:2006sp,Leitner:2006ww,Leitner:2008wx,Leitner:2009ec,Leitner:2009de,Lalakulich:2010ss,Leitner:2010jv,Leitner:2010kp,Lalakulich:2011ne,Lalakulich:2012ac}.

 Neutrinos are the secondary decay products of pions or kaons produced usually in a high-energy proton-nucleus collision. (K2K: 12.9 GeV p on Al and MiniBooNE: 8.9 GeV p on Be) The energy-distributions of LBL neutrino beams are thus quite broad (several hundred MeV to several GeV). Dedicated experiments, such as HARP \cite{harp} and NA61/SHINE \cite{na61}, measure the pion and kaon production rates and thus provide crucial input for the neutrino beam simulations. For a typical nuclear physicist LBL experiments are quite unbelievable: the beam at its source is already about 0.5 m wide, at the target it can reach out over several hundred meters. The 'beamlines' are long, up to several hundred kilometers, the composition of the beam is known only up to a certain extent and, most important, the energy of the beam is distributed in a wide range and is not really well determined. Taken all of this together with the fact that typical cross sections for interaction with matter are of the order $10^{-38}$ cm$^2$ it is even more remarkable that these experiments have actually given us the first look into physics beyond the Standard model. Since neutrino oscillations have been verified in different experiments~\cite{Fukuda:1998mi,Ahmad:2001an} there is now no longer any doubt that neutrinos have a non-zero, although quite small, rest mass. At present there are several so-called long-baseline (LBL) experiments running with neutrino energies of a few hundred MeV up to several GeV~\cite{minos,nova,k2k,t2k,opera}. All these experiments aim for a more refined determination of the neutrino mass-differences and the oscillation parameters. They also look for fundamental effects such as CP-violation in electroweak interactions, the possible existence of so-called sterile neutrinos and cross section measurements \cite{miniboone,sciboone,Drakoulakos:2004gn}.

For the determination of the mass-differences the neutrino energy has to be known as is evident from the simplified (two flavors only) neutrino oscillation formula
\begin{equation}
P(\nu_{\mu}\rightarrow \nu_{e})= \sin^2 (2  \theta)  \sin^2
\left( \frac{\Delta m^2 L }{4 E_{\nu}} \right) ~
\end{equation}
which gives the probability that a muon neutrino $\nu_\mu$ converts into an electron neutrino $\nu_e$ over the flight path $L$.
Here $\Delta m^2$ is the difference of the squared masses of the two neutrino flavors and $\theta$ is the mixing angle. It is seen that with known distance $L$ between source and detector the difference of the squared masses is directly proportional to the neutrino energy $E_\nu$. The neutrino energy is thus a crucial quantity which has to be known on an event-by-event basis. Due to the production mechanism, however, only its distribution in the neutrino beam is known. The neutrino energy thus has to be reconstructed from the measured energies of the outgoing lepton (for charged-current (CC) experiments) and/or the hadrons in the final state of the reaction. For a quasielastic scattering (QE) reaction on a free nucleon at rest the incoming neutrino energy is directly linked to the kinematics of the outgoing lepton and is thus known when lepton angle and energy are measured.

 Due to practical considerations (difficulties in handling hydrogen targets and to increase the count rates) all of these new experiments use nuclear targets (e.g.\ O, C or Fe). Nuclear effects on the measured inclusive cross sections have been analyzed with the help of Fermi gas models that contain the Fermi motion of the nucleons inside the nuclear targets, Pauli-blocking and also -- roughly -- the nuclear binding~\cite{Smith:1972xh,Donnelly:1978tz,Horowitz:1993rj,Barbaro:1996vd,Alberico:1997vh,Alberico:1997rm}. In addition, nucleons inside nuclei can be collision-broadened. For example, Benhar~\cite{Benhar:2005dj,Benhar:2006nr,Benhar:2010nx} calculates their spectral function within a state-of-the-art nuclear many-body theory. Finally, using RPA also corrections of the nucleonic vertex have been calculated~\cite{Singh:1992dc,Marteau:1999kt,Nieves:2004wx,Martini:2009uj,Martini:2010ex,Nieves:2011pp}. Other approaches go beyond the Fermi gas model and in particular a plane-wave approximation and calculate QE scattering using more realistic wave functions~\cite{Alberico:1997vh,Alberico:1997rm,Jachowicz:1998fn,Meucci:2004ip,Martinez:2005xe,vanderVentel:2005ke,Meucci:2006ir,Meucci:2011vd}. Also the so-called scaling analysis of inclusive electron scattering has been used for a prediction of neutrino scattering~\cite{Amaro:2004bs,Amaro:2010sd,Amaro:2011qb}. Common to all of these theoretical attempts is that they concentrate on QE alone. At energies of a few hundred MeV (as in the MiniBooNE and K2K experiments) QE indeed dominates the reaction process. However, even then there are higher-energy tails in the energy distributions of the neutrinos so that pion-production events are always entangled with QE. All the calculations mentioned are thus not directly comparable with 'raw' experimental data which always contain some admixture of pion production events. Free QE kinematics are then not directly applicable.

Semi-inclusive processes, in which a nucleon is knocked out from the nucleus, seem to offer direct access to QE events. For these processes one needs obviously a very good description of FSI. Often, these FSI are completely neglected~\cite{Horowitz:1993rj,Barbaro:1996vd,vanderVentel:2005ke} or they are being treated only as absorptive interactions, in the framework of an optical model or the Glauber approximation~\cite{Alberico:1997vh,Alberico:1997rm,Meucci:2004ip,Martinez:2005xe,Meucci:2006ir,vanderVentel:2003km}. In doing so one neglects that collisions of the initially knocked-on nucleons with other nucleons in the nuclear target can lead to energy loss, change of direction, charge transfer or to multiple nucleon knock-out. These processes can only be described by Monte Carlo models~\cite{Nieves:2005rq,Buss:2011mx}. Contributions to nucleon knock-out come also from resonance excitations. For example, a $\Delta$ resonance can first be excited which then decays by a collision with another nucleon thus leading to two outgoing nucleons which resemble a QE knock-out event. These 'fake QE' events are very important for the experiments since they have a big influence on the energy reconstruction as will be demonstrated later on in this article. A quantitative understanding of these events is only possible within coupled channel models that contain all the various interaction mechanisms.

Pion production can not only influence the energy reconstruction, but is essential also for the correct interpretation of neutrino oscillation experiments. In particular neutral pions play an essential role in the flavor identification which is essential since every neutrino beam always contains admixtures of less dominant flavors. For example, a $\nu_\mu$-produced $\pi^0$, which may be produced by FSI, can decay to $2 \gamma$ and thus simulate a $\nu_e$ event in the detector. Up to neutrino energies of about 1.5 GeV pion production proceeds mainly through the excitation and subsequent decay of the $\Delta$ resonance. With increasing energy also higher resonances and non-resonant background can contribute~\cite{Fogli:1979cz,Paschos:2003qr,Lalakulich:2010ss}. Singh et al.~\cite{Singh:1998ha} have shown that beside medium effects such as Fermi-motion, Pauli-blocking and the binding of the nucleons in their potential also more complex in-medium effects such as the collisional broadening of nucleons may influence the width of the $\Delta$ and through it the absorption cross section. Before we started our own work in this field, semi-inclusive observables such as pion spectra had been treated only within the old ANP model~\cite{Paschos:2000be}.

At high neutrino energies finally deep inelastic scattering (DIS) dominates. Even though nuclear effects have been discussed in this energy region for the NuTeV Experiment~\cite{Zeller:2001hh} ($\sin^2 \theta_W$-anomaly) most analyses assume that here the nucleus just acts as an ensemble of free nucleons~\cite{Conrad:1997ne}. However, just as there is an EMC effect for electron- and mu-induced reactions on nuclei a similar effect should be present for neutrinos as well. This is still a field of ongoing research; very recent analyzes do not seem to be compatible with the electron-data \cite{Schienbein:2007fs,Kovarik:2010uv}.


For a reliable interpretation of the ongoing neutrino experiments one needs a model that describes both the initial neutrino-nucleus interaction \emph{and} the final state interaction of the produced particles. Since, because of the always present broad energy distributions in neutrino beams, any experiment inherently averages over different reaction types, the model has to be reliable and well tested in all three mentioned energy regimes (QE, pion production, DIS). Since none of the works mentioned earlier could achieve this goal, special event generators were -- quite independently from other nuclear physics events generators -- developed for the simulation of the various neutrino experiments (e.g.\ ~NUANCE~\cite{Casper:2002sd}, NEUGEN~\cite{Gallagher:2002sf}, NEUT~\cite{Hayato:2002sd}) and GENIE~\cite{GENIE}). Although these different generators are quite different from each other, common to all of them is that they all rely on Monte Carlo simulations of the reaction process. These codes contain contributions from quasielastic scattering as well as from inelastic processes; for the resonance excitations the model of Rein and Sehgal~\cite{Rein:1980wg} is widely used. They all use a Fermi-gas model for the nucleons in the target nucleus, usually with constant Fermi-momentum and constant binding energy. Final state interactions are then described by an intranuclear cascade, where the scattering and absorption cross sections are adjusted to the special targets used. Very few predictions of these models have been compared to data for other reaction types. This is particularly disturbing since there exist extensive data for photonuclear and electronuclear experiments on nuclear targets in the relevant energy regime. Only recently, tests of selected aspects of FSI, i.e.\ pion absorption, have been started \cite{Dytman:2011zz}. Furthermore, the nuclear physics input of these model is often quite doubtful. For example, the Rein-Sehgal formfactors fail when applied to electroproduction experiments \cite{Leitner:2008fg} and in the analysis of the MiniBooNE data even the strength of the Pauli-principle was adjusted to QE data \cite{AguilarArevalo:2010zc}.

\section{The GiBUU model}

Over the last 25 years we have gained experience in the development of theories, numerical methods and codes for the simulation of complete events in nuclear reactions. These methods all are based on the Boltzmann-Uehling-Uhlenbeck (BUU) equation, which differs from the mentioned neutrino event generators by a consistent treatment of self-energies of all participating particles. Fermi-motion in the target nucleus is described by a local Thomas-Fermi approximation and is thus closer to empirical distributions than this is normally the case in event generators; the occupation in coordinate space follows empirical density distributions. A recent review summarizes some of our work with the GiBUU implementation of this method \cite{Buss:2011mx} and gives many more details. GiBUU essentially factorizes a reaction on a nuclear target into a first, initial interaction (sufficient for inclusive cross sections) and a detailed treatment of FSI (needed for a reliable description of semi-inclusive processes). The latter is the particular strength of GiBUU.

With this method we have initially described the production of quite different particles (photons, dileptons, pions, kaons, vector mesons etc.) in heavy ion reactions~\cite{Blaettel:1993uz,Cassing:1990dr}. Later on, we have generalized these model studies to more elementary projectiles and have investigated proton-induced~\cite{Bratkovskaya:2000mb}, pion-induced~\cite{Weidmann:1997vj,Effenberger:1999nn},
photon-induced~\cite{Hombach:1994gb,Effenberger:1996rc,Effenberger:1999ay,Buss:2006yk,Muhlich:2004zj}
and electron-induced~\cite{Lehr:1999zr,BussDiss:2008} particle production on nuclei over a wide energy range, using the same theoretical methods \emph{and} code for all these different reactions. In particular, the electron-induced reactions are directly relevant for neutrino-induced reactions as they provide a necessary testing ground for the neutrino event generators since the initial kinematics is similar and the primary reaction amplitudes are closely related. In addition, the hadronic FSI are obviously the same. The electron-induced reactions with their large data base are thus ideally suited for a test of the treatment of FSI.

Indeed, our studies of photon-induced reactions~\cite{Krusche:2004uw,Buss:2006vh} have shown that a detailed, quantitatively reliable description of
the experimental data is possible only if the FSI of the produced hadrons is described realistically. In particular, the effects of coupled channels
must be taken into account which can lead to sidefeeding of the channel under study. In other words, the particle that the detector finally -- after its traverse through the nucleus -- sees must not necessarily be the same as the one originally produced in the first interaction. We have shown in studies of electroproduction \cite{Lehr:1999zr} that the importance of such secondary production processes increases with the squared momentum transfer $Q^2$. We have also performed extended studies of pion mean free paths and their absorption in nuclear targets~\cite{Buss:2006vh}. Very sensitive to details of the $\pi N$ interaction are charge exchange processes. Their reliable description is thus an important test of the reliability of our model for the FSI~\cite{Buss:2006yk}.

The correct description of the imaginary part of the self-energies of the produced particles, i.e.\ their absorption through interactions with the nucleus, is very important in all these calculations. Particles can get 'collision-broadened' in the nuclear medium, i.e.\ the imaginary parts of their self-energies can become large \cite{Lehr:2001qy,Benhar:2006wy}. It is then a challenge to describe the transport of these particles so that they attain their free self-energies when they reach the nuclear surface. This is one of the major developments of transport theories in the last decade. We are now in a position to transport collision-broadened hadrons consistently and theoretically well founded (so-called off-shell transport) \cite{Effenberger:1999ay,Leupold:2000ma,Cassing:1999wx,Buss:2011mx}. In this point we thus go beyond the simpler approaches in which e.g.\ nucleon resonances are not transported at all, but instead decay immediately; this corresponds to a local approximation for the resonance propagators. This point is essential in particular in light and medium-heavy nuclei where the change of the density-profile of the target nucleus over the propagation-length of the resonance plays an important role.

At invariant masses higher than about 2 GeV the discrete resonance structure disappears and here the region of quark-hadron duality sets in. We, therefore, employ in this regime well-tested methods of pQCD in which the total inclusive cross section is described with the help of parton distribution functions and the non-perturbative formation of hadrons is encoded in fragmentation functions. The latter we obtain from the JETSET (Lund) model which is implemented in the code PYTHIA \cite{Sjostrand:2006za}. Also this high-energy part we have tested extensively. In particular, we have investigated the hadron formation and attenuation in lepton-induced reactions as measured in the HERMES and EMC experiments~\cite{Gallmeister:2007an}. Data from both experiments range from 12 GeV (HERMES) up to 280 GeV (EMC) and can be described simultaneously by our model. These descriptions also offer access to the so-called formation time of hadrons during which FSI are suppressed; for details see our discussions in ~\cite{Gallmeister:2007an}.

Based on all these developments we have extended the GiBUU model over the last few years also to neutrino-induced reactions \cite{Leitner:2006sp,Leitner:2006ww}.
The GiBUU model differs from the standard neutrino event generators not only by its theoretical foundation. It is, furthermore, the only available method that has been tested in comparisons with broad data bases from quite different experiments~\cite{gibuu}.  Nuclear many-body physics is complex and a realistic treatment of FSI requires correspondingly a complex model. This is reflected in the quite sizeable GiBUU code which is available for download from \cite{gibuu}. There also a description of the numerical implementation can be found; further details of the GiBUU model are contained in the review \cite{Buss:2011mx}.

\section{Results}
As we have discussed already in the introduction one of the essential ingredients for any extraction of the neutrino masses from oscillation experiments is the neutrino energy. This energy is not known a priori in present-days experiments, because neutrino beams are quite broad in energy due to their production mechanisms. While at higher energies calorimetric methods may play a role, at lower energies (a few hundred MeV to a few GeV) quasielastic (QE) scattering has been used to determine the incoming neutrino energy on an event-by-event basis. This method relies on an identification of the reaction mechanism (interaction of the neutrino with a single nucleon). It also relies on the use of quasifree kinematics that describes neutrino scattering on a single free nucleon at rest, thus neglecting  any Fermi-motion effects; binding is taken
into account only by a constant removal energy.

\subsection{The QE-Pion entanglement}

 The experimental identification of a special reaction type, in this case QE, is, however, anything but trivial in an experiment with a broad energy distribution in the incoming beam. There is necessarily a mixture of different reaction mechanisms. Up to energies of about 1.5 GeV these are mainly QE and pion production. The experiments have thus developed strategies to isolate the QE process. In Cerenkov detectors this is done by requiring only the outgoing lepton and the absence of any pions. In so-called tracking detectors 1 muon, 1 proton and 0 pions are required for a QE event. We call these events 'QE-like'.

\begin{figure}[tbp]
  \centering
  \includegraphics[scale=\plotscale]{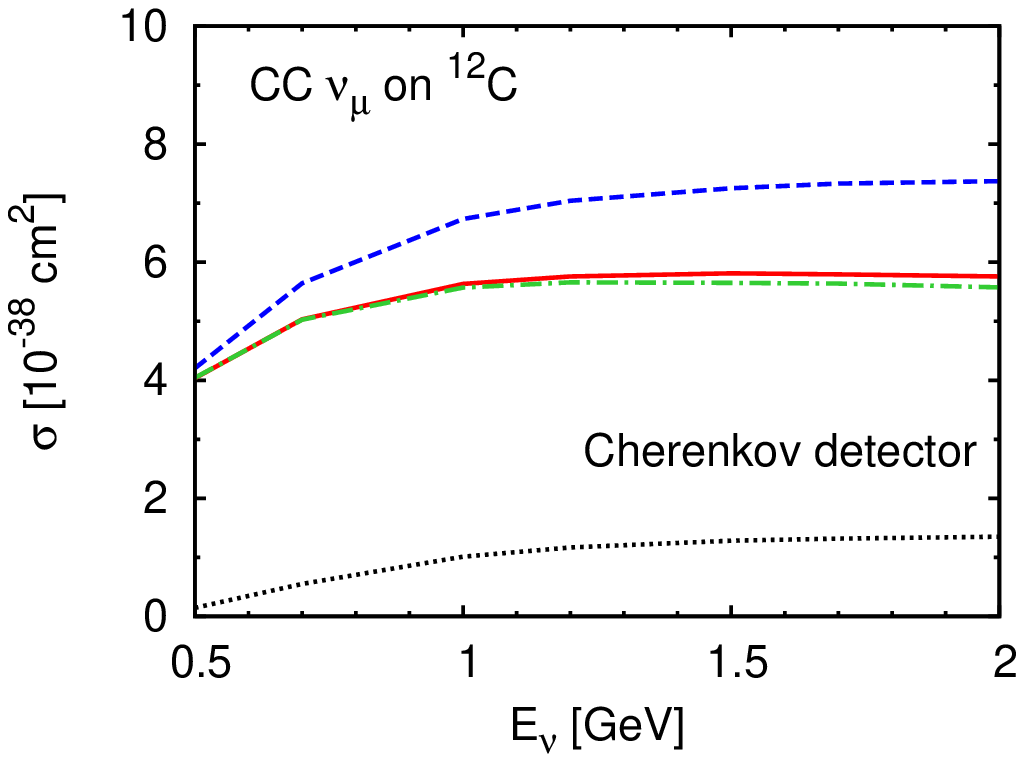}
  \includegraphics[scale=\plotscale]{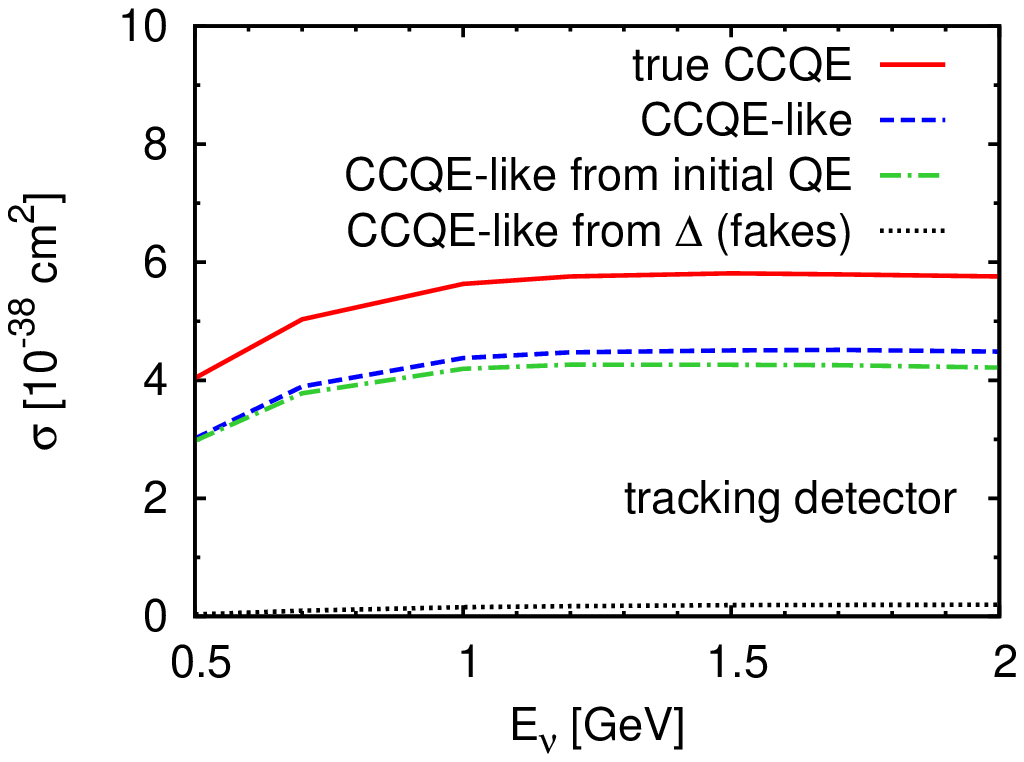}
  \caption{(Color online) Total QE cross section on \carbon{} (solid lines) compared to
    different methods on how to identify CCQE-like events in experiments (dashed lines).
    The left panel shows the method commonly applied in Cherenkov detectors; the right
    panel shows the tracking-detector method as described in the text. The contributions to the
    CCQE-like events are also classified [CCQE-like from initial QE (dash-dotted) and from
    initial $\Delta$ (dotted lines)]. Experimental detection thresholds are not taken into
    account ( from \cite{Leitner:2010kp}).
    \label{fig:QEmethods}}
\end{figure}

Fig.\ \ref{fig:QEmethods} shows the cross sections for scattering on $^{12}C$ as a function of neutrino energy both for the Cherenkov ($\mu,\, 0\pi$) and the tracking detector ($\mu,\, 1p, 0\pi$) identification methods \cite{Leitner:2010kp} for QE-like events. The solid curve in both cases shows the true CCQE cross section whereas the dashed line gives the QE-like cross section. It is seen that the Cherenkov detector identification method leads to a cross section that is about 20\% \emph{higher} than the true value; the surplus is due to primary resonance or pion production and subsequent FSI leading to pionless final states. For determination of the axial mass or for the reconstruction of the incoming neutrino energy this artificial surplus has to be removed with the help of an event generator. The opposite is the case for the tracking detector. Here the QE-like cross section is about 20\% \emph{lower} than the true value; this is because, first, more than one proton may be knocked-out of the nucleus and these events are then not counted as QE-like and, second, secondary neutrons are not detectable. However, the QE-like sample is very 'clean' in that it contains nearly only original QE events.

A further complication arises because detectors are not perfect but have acceptance thresholds.
Their effect has been discussed in detail in \cite{Leitner:2010jv}. There it was shown that for realistic detector thresholds only about 50 - 70\% of all events are observed and that this deficit also affects the energy reconstruction.

\subsection{The QE puzzle}

In theoretical calculations the QE cross section is determined by an interplay of vector and axial couplings with their corresponding form factors. The vector couplings can be rather well determined from electron scattering experiments on the nucleon that work at a fixed energy and permit to determine the relevant kinematic parameters, i.e.\ energy- and momentum-transfer, in each event. The corresponding form factors have been shown to have a complicated, non-dipole form \cite{Arrington:2006zm}. For the axial couplings the situation is less well determined. Here the data come from electro-pion production and older neutrino data on the nucleon or deuterium with large uncertainties. They have been analyzed by making a dipole ansatz and then extracting the axial mass from a fit to data. The world average for the axial mass parameter is found to be $M_A = 1.026 \GeV$ \cite{Bernard:2001rs}.

\begin{figure}[hbt]
\includegraphics*[trim = 0mm 0mm 0mm 70mm, width = 13cm]{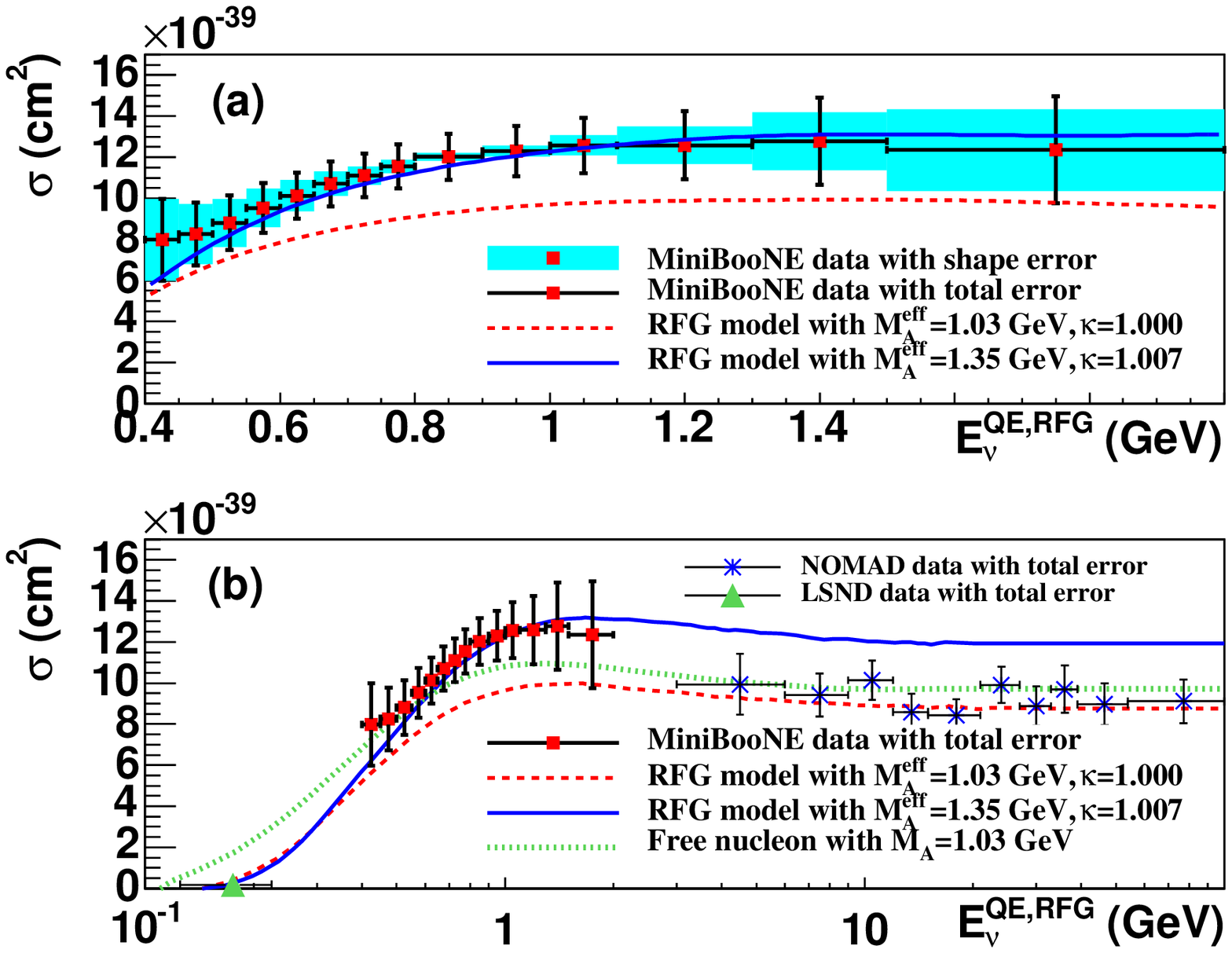}
\caption{(Color online) Flux-unfolded MiniBooNE CCQE
cross section per neutron as a function of reconstructed neutrino energy. The cross sections are shown
along with results from the LSND \cite{Auerbach:2002iy} and NOMAD \cite{Lyubushkin:2008pe} experiments. Also shown are predictions from the NUANCE simulation for a relativistic Fermi gas model with two different parameter variations and for scattering from free nucleons with the world average $M_A$ value ($\kappa$ is a parameter influencing the strength of the Pauli-principle).
Figure from~\protect\cite{AguilarArevalo:2010zc}.}
\label{fig:MBsigma}
\end{figure}

It came, therefore, as a surprise when the Mini Booster Neutrino Experiment (MiniBooNE) at Fermilab \cite{miniboone}, which uses a big container filled with oil as a Cherenkov counter, published its results on QE scattering. The analyses of both charged current (CC) and neutral current (NC) high-statistics QE events showed a clear excess of cross section over that expected for QE scattering from a Fermi-gas model \cite{:2007ru,AguilarArevalo:2010zc,AguilarArevalo:2010cx}; see Fig.\ \ref{fig:MBsigma}.
A similar result had been obtained by the K2K experiment \cite{Gran:2006jn} that worked with a neutrino flux peaked at a slightly higher neutrino energy ($\approx 1 \GeV$) than the MiniBooNE experiment ($\approx 0.7 \GeV$). In both cases the flux distributions are rather broad and have a considerable overlap. On the contrary, the NOMAD experiment working at significantly higher energies (between about 5 and $100\GeV$) observed no such excess of measured over expected quasielastic cross section \cite{Lyubushkin:2008pe}.
Both the MiniBooNE and the K2K experiments could obtain good fits to their data\footnote{It must be noted that neither the cross section nor the energy in Fig.\ \ref{fig:MBsigma} have been directly measured; both involve some model dependence.} in the Fermi-gas model only when the axial mass was considerably increased to $M_A = 1.23 \GeV$ \cite{:2007ru} or even $M_A = 1.35 \GeV$ \cite{AguilarArevalo:2010zc}. The sizeable increase in the axial mass needed to describe the data cannot be ascribed to deficiencies in the Fermi gas model alone. Indeed, Benhar et al.\ showed that a model based on state-of-the-art nucleon spectral functions required an even larger axial mass ($M_A = 1.6 \GeV$) for a fit of the differential data \cite{Benhar:2010nx}.

\subsection{The pion puzzle}

Recently the MiniBooNE and K2K collaborations have published data on
charged~\cite{AguilarArevalo:2010bm} and neutral~\cite{AguilarArevalo:2010xt} pion production
in CC neutrino scattering. Generally, these experiments  report cross sections which are noticeably higher
than those expected from any conventional theoretical approach \cite{Dytman:2009zzb,Leitner:2009de}.

\begin{figure}[hbt]
\centering
\includegraphics[width=0.8\linewidth]{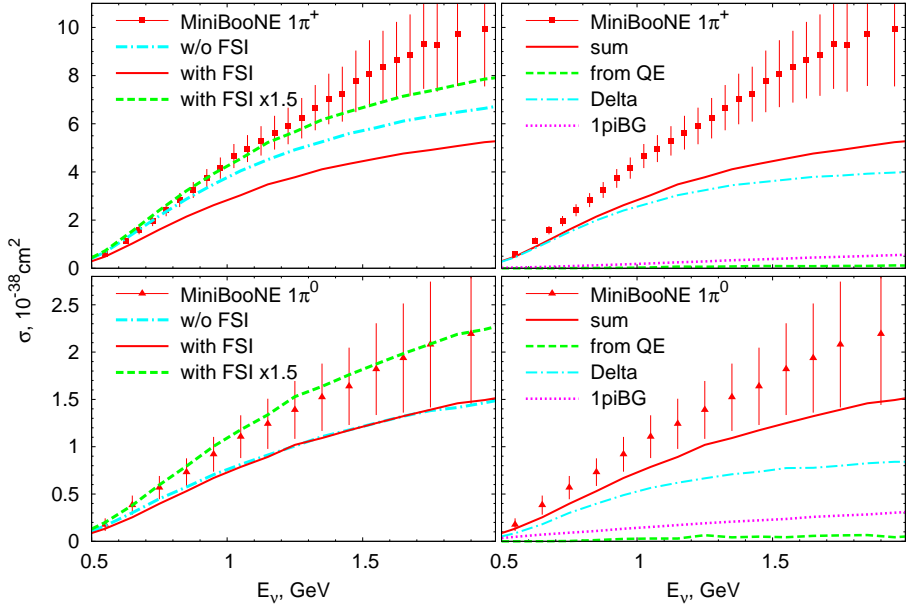}
\caption{(Color online) Integrated cross section for MiniBooNE $1\pi^+$(upper panel) and
$1\pi^0$(lower panel) CC production versus reconstructed neutrino energy.
Data are from~\protect\cite{AguilarArevalo:2010bm,AguilarArevalo:2010xt}.
The panels on the right show the composition of the calculated cross sections. Figure from \cite{Lalakulich:2011ne}}
\label{fig:MB-lepton-Enu-QEDelta}
\end{figure}

As an illustration we present our calculations on a $\mathrm{CH}_2$ target and compare the results with the data
from the MiniBooNE experiment \cite{AguilarArevalo:2010bm,AguilarArevalo:2010xt}.
Integrated cross sections versus neutrino energy for the charged current $1\pi^+$ and $1\pi^0$ production
are  shown in Fig.~\ref{fig:MB-lepton-Enu-QEDelta}.
As in the MiniBooNE experiment, the \onepion events are defined as ``observable \onepion production'',
i.e. events  with one pion of a given charge and no other pions in the final state,
regardless of which particles  were produced in the initial neutrino vertex.

The left upper panel in Fig.~\ref{fig:MB-lepton-Enu-QEDelta} shows the results for $\pi^+$ production.
Comparison of the curves with and without FSI shows that the FSI do not change the energy-dependence of the curves.
The curve with FSI (solid line) lies 20\% below the curve without FSI (dash-dotted line).
This reduction is mainly caused by the $\Delta$ absorption through $N \Delta \to NN$ scattering.
Charge exchange $\pi^+ n \to \pi^0 p$ also depletes the dominant $1\pi^+$ channel.
The side feeding from the reverse
process gives minor relative contribution because the initial $\pi^0$ production cross section is around 5 times lower.
For the 1$\pi^0$ cross section in the left lower panel the same side feeding processes
increase the cross section. This is, however, compensated by absorption, charge exchange to the $\pi^-$ channel through
$\pi^0 n \to \pi^- p$ and other channels such as $\pi N \to \Sigma K, \Lambda K$. The overall effect is nearly the same
cross section with and without FSI.

The right upper panel shows the origin of the $1\pi^+$ events. Most of them (dash-dotted line)
come from initial  $\Delta$ resonance production and its following decay. Some  events (dotted line) are
background ones. A minor amount comes from the initial QE vertex (long-dashed line),
which is only possible due to FSI, when the outgoing proton is rescattered.
Here the main contribution is from  the $p N \to N' \Delta \to N' N^{''} \pi$ reaction.
The right lower panel shows the origin of the $1\pi^0$ events. Here the background processes and the FSI
play an even bigger role.

We now turn to a discussion of a comparison with experiment. Already the curves \emph{without}
any FSI  lie considerably ($\approx 25$\%) \emph{below} the data,
those with FSI included (solid line) are nearly by a factor of about 1.5 (at $1 \GeV$)
below the experimental data;
at higher neutrino energies these discrepancies become even larger since the experimental cross
sections rise more steeply with energy than the  calculated pion cross sections.
At 2 GeV the discrepancy between the results with FSI and the data amounts to a factor of $\approx 2$.
For 1$\pi^0$ production  one observes the similar result: the data are considerably higher than our calculated values.

\begin{figure}[!hbt]
\begin{minipage}[c]{0.48\textwidth}
\includegraphics[width=\textwidth]{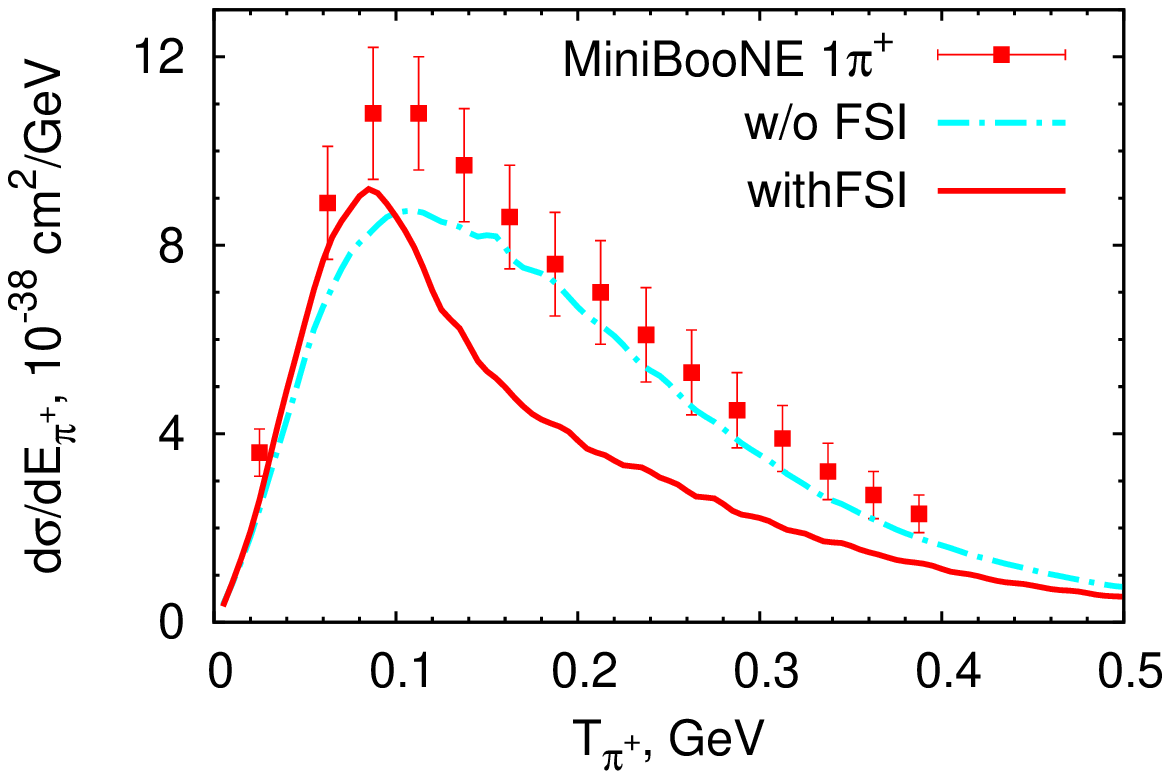}
\end{minipage}
\hfill
\begin{minipage}[c]{0.48\textwidth}
\includegraphics[width=\textwidth]{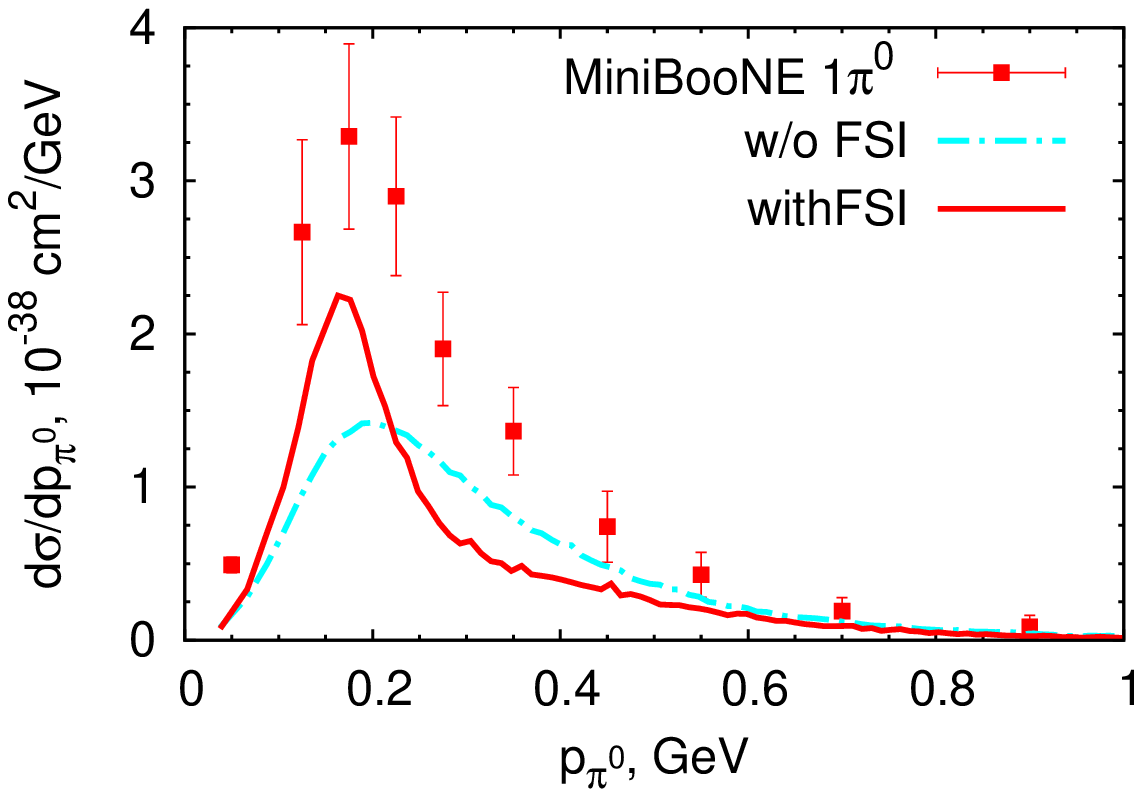}
\end{minipage}
\caption{Distribution of (left panel) the outgoing $\pi^+$ in their kinetic energy; (right panel)
the outgoing $\pi^0$ in their absolute value of the 3-momentum  for the MiniBooNE CC neutrino scattering.
Data are from~\protect\cite{AguilarArevalo:2010bm,AguilarArevalo:2010xt}, figure from \cite{Lalakulich:2011ne}.}
\label{fig:MB-pion-dTkin}
\end{figure}

Fig.~\ref{fig:MB-pion-dTkin} shows our calculations for the kinetic energy distribution of the outgoing pions.
As for the other distributions, our calculations with FSI are lower than the experimental data by a factor of
$1.6 - 2$.

A distinctive feature of our calculations is the result that FSI significantly
change the pion spectra.
The lowering of the cross section for $T_{\pi} > 0.12 \GeV$
 is a direct consequence of the
pion absorption through $\pi N \to \Delta$ with the following $\Delta N \to N N$.
Pion elastic scattering in the FSI also  decreases the pion energy,
thus depleting spectra at high energies
and accumulating strength at low energies.
For $\pi^0$ production an additional increase of the cross section at lower energies
comes from the side feeding from the dominant $\pi^+$ channel as discussed above.
The change of the shape of the spectra is similar to that calculated for
neutral current 1$\pi^0$ production in \cite{Leitner:2008wx}.

The predicted shape of the pion spectra is due to well-known $\pi-N-\Delta$ dynamics in the medium is the same as
that observed experimentally in $(\gamma, \pi^0)$
production on nuclear targets \cite{Krusche:2004uw,Mertens:2008np}.
Its absence in the neutrino data is, therefore, hard to understand \cite{Leitner:2009de,AlvarezRuso:2010ia}.

A source of uncertainties in these comparisons is the
determination of the neutrino energy which is done using quasi-free kinematics
for on-shell Delta production on a nucleon at rest. One more point is that the experimental data
are presented after being corrected for the finite detection thresholds for muons,
pions and nucleons. This may introduce an additional dependence on a particular neutrino
event generator used in a given experiment \cite{Leitner:2010kp}.

\subsection{Energy Reconstruction}

\begin{figure}[th]
  \centering
  \includegraphics[scale=\plotscale]{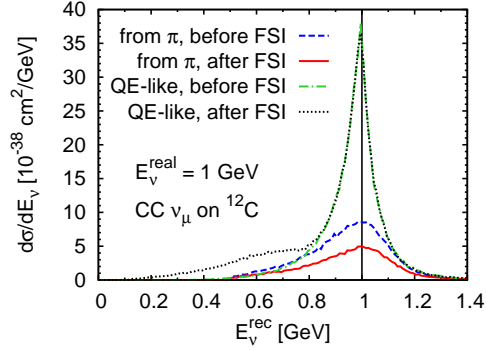}
  \caption{(Color online) Distribution of the reconstructed neutrino energy using quasifree kinematics
    on a nucleon at rest for QE and for static $\Delta$ excitation for $E_\nu^\text{real}=1$ GeV. Shown is the reconstruction based on the CCQE-like sample (before and after fsi and Cherenkov assumptions) and based
    on the CC$1\pi^+$ sample (before and after fsi). Notice that the dash-dotted and dotted curves partially overlap (from \cite{Leitner:2010kp}).
    \label{fig:energyrec_ratio}}
\end{figure}

We now discuss the quality of the energy reconstruction which is based on applying quasifree kinematics of true QE scattering to QE-like events and neglecting Fermi motion \cite{Abe:2011sj}. Fig.\ \ref{fig:energyrec_ratio} shows the distribution of reconstructed energies, obtained in a GiBUU simulation, for a fixed incoming neutrino energy of 1 GeV. The distribution is clearly affected by the two effects discussed above: Fermi motion leads to a broadening of the neutrino energy around the incoming energy; this accounts for the broadened peak at 1 GeV. Its width of about 16\% is determined by Fermi motion alone and is thus always present; this defines the lower limit for any energy reconstruction via QE scattering. In addition the reconstructed energy distribution exhibits a clear bump at lower energies; this is caused by an initial pion production. The bump does depend on the pion detection threshold and amounts to about 15\% of the true QE peak height at a realistic detection threshold of about 100 MeV pion kinetic energy. Taking this lower-energy bump into account raises the rms energy-width to about 22 \% (for a more detailed discussion see \cite{Leitner:2010kp}). A qualitatively similar result has also been obtained by Tanaka (see Fig.\ 3 in \cite{Tanaka:2009zzb}) using the GENIE event generator; there the lower-energy bump is even more pronounced.

\begin{figure}[htb]
  \centering
  \includegraphics[scale=\plotscale]{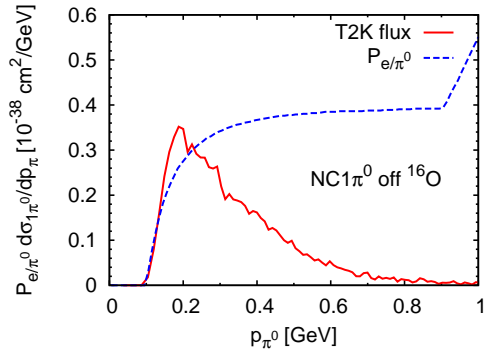}
  \caption{(Color online) NC induced single-$\pi^0$ production cross sections on \oxygen{}
    averaged over the T2K flux and multiplied by the misidentification
    probability (dashed line, taken from \refcite{Okamura:2010zz}) as a function of the
    pion momentum (from \cite{Leitner:2010jv}).
    \label{fig:T2K_NC_misid}}
\end{figure}

The T2K experiment has recently reported the first experimental observation of electron neutrino appearance from a muon neutrino beam \cite{Abe:2011sj}. At the flux maximum the energy of is $\approx 600$ MeV. At this energy the error in the reconstructed energy, according to Table I in \cite{Leitner:2010kp}, amounts to about 21\% for the Cherenkov detector. Fig.\ \ref{fig:QEmethods} shows that at 600 MeV the QE events are fairly clean, but that the admixture of pion production events rises significantly, already up to 1 GeV, reflecting the pion production threshold.

A crucial problem in any such experiment with Cherenkov counters is that the decay photons
of neutral pions can be misidentified as electrons. We therefore, investigate in Fig.\
\ref{fig:T2K_NC_misid} the probability that a misidentified $\pi^0$ is counted as a
$\nu_e$ appearance event. The probability that a $\pi^0$ cannot be distinguished from
$e^\pm$ is given by the dashed line (taken from Fig.~2 of \refcite{Okamura:2010zz}). The solid line shows the
weighted cross section averaged over the T2K flux and calculated using the elementary ANL
data as input. The total cross section for misidentified events is now
$0.09\,\cdot\,10^{-38}\, \text{cm}^2$ and thus 26\% of the true pion events.

\subsection{Many-body interactions}

All the results discussed so far rely on the impulse approximation. It has, however, recently been pointed out that the identification method of the MiniBooNE experiment allows for a significant amount of non-QE many-body excitations in the QE-like cross section \cite{Martini:2009uj,Martini:2011wp,Nieves:2011pp,Nieves:2011yp}. These excitations are connected with more than one outgoing nucleon in the initial, primary interaction and can be caused by the interaction of the incoming neutrino with more than one nucleon. That such reactions contribute to the inclusive electron scattering cross sections on nuclei is quite well known since about 20 years when it was shown that such events contribute significantly in the so-called dip region between the QE peak and the $\Delta$ resonance. For a more extended discussion of the literature see \cite{Lalakulich:2012ac}.

 These so-called 2 particle - 2 hole (2p-2h) effects can indeed account for the disagreement between the data in Fig.\ \ref{fig:MBsigma} and the calculations using the world average value of about 1 GeV for the axial mass and are not incorporated in the energy reconstruction discussed in the last section. Since for the 2p-2h excitations the quasifree kinematics formulas used by experiment for the energy reconstruction do not apply, the actual energy uncertainty may even be larger than discussed before.
There are two features of of these 2p-2h contributions that affect the energy reconstruction.  First, at forward angles, where the cross section is largest, the 2p-2h contributions are largest at small $T_\mu$, below the QE peak. When analyzing such events with the help of the one-body expression this leads to a lower reconstructed neutrino energy than the true one. Second, while QE scattering is strongly forward peaked, the 2p-2h events are fairly flat (within a factor of 2) in lepton angle (see \cite{Lalakulich:2012ac}). The relatively strong  yield at backward angles will lead to a larger neutrino energy, in particular for intermediate muon energies. Since both effects are present we expect a fairly flat behavior of the 2p-2h contribution to the energy reconstruction.
\begin{figure}[htb]
\centering
\includegraphics[scale=\plotscale]{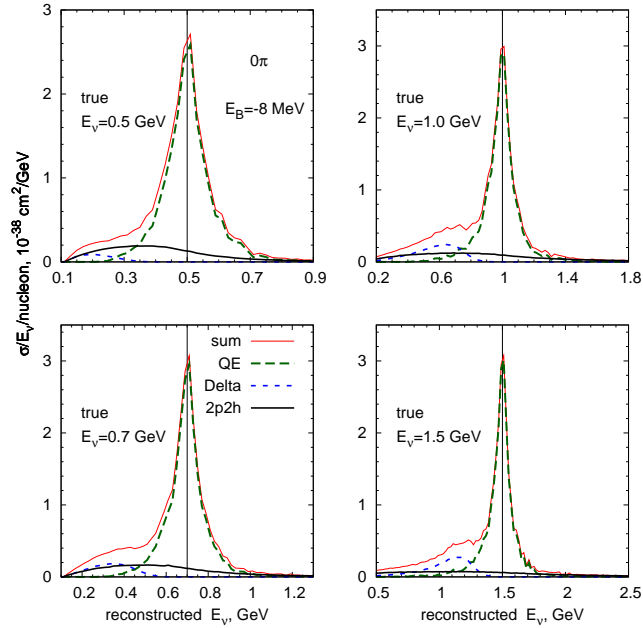}
\caption{(color online) Cross section per energy per nucleon for QE-like processes as defined by the MiniBooNE experiment (0 pions and any number of nucleons in the final states) versus the reconstructed neutrino energy for the true energy of the incoming neutrino of $0.5, \ 0.7, \ 1.0$ and $1.5 \GeV$. Figure from \cite{Lalakulich:2012ac}.
}
\label{fig:enu-reconstruction-fixedEnu-nopi-ME8}
\end{figure}

We have recently implemented the 2p-2h excitations into GiBUU \cite{Lalakulich:2012ac}.
In Fig.~\ref{fig:enu-reconstruction-fixedEnu-nopi-ME8} we plot the distribution of the reconstructed neutrino energy obtained using the MiniBooNE reconstruction method with  $E_B=-8 \MeV$ (which is a typical binding energy in the GiBUU code, as opposed to the value $-34 \MeV$ used by the MiniBooNE) for the fixed true neutrino energies $E_\nu^\mathrm{true}=0.5, \ 0.7, \ 1.0$ and $1.5\GeV$. The main contribution to all QE-like events is given by the true QE events, which shows a prominent peak around the real energy. The peak is approximately symmetric and has a width of about $0.1\GeV$: this broadening is caused by Fermi motion. For $\Delta-$induced events (as already discussed in \cite{Leitner:2010kp} and in the previous section) the distribution is not symmetric, with a broad peak at lower energies. For a more detailed discussion of these effects see \cite{Leitner:2010jv,Leitner:2010kp}.

For all energies the broad 2p-2h contribution is peaked below the true energy. For the lower neutrino energies of $0.5$ and $0.7\GeV$ most of the total distortion is caused by 2p-2h processes because at this low energy there is only little $\Delta$ excitation. For the true neutrino energies of 1.0 and $1.5\GeV$, on the other hand, the contributions of 2p-2h events are comparably small because here $\Delta$ excitation plays a major role and because their strength is distributed over the whole energy range from $0.2$ to $1.8\GeV$, with a flat maximum around 1/2 of the true energy. Both the 2p-2h effects as well as the $\Delta$ excitation lead to a shift of the reconstructed energy towards smaller values, or, vice versa, for a given reconstructed energy the true energy always lies higher than the reconstructed one. The effect is most pronounced at lower true energies and becomes smaller at higher energies. Both of these results agree with a recent analysis by  Martini et al.\ \cite{Martini:2012ff}.

\begin{figure}[hbt]
\centering
\begin{minipage}[c]{0.48\textwidth}
\includegraphics[scale={\plotscale}]{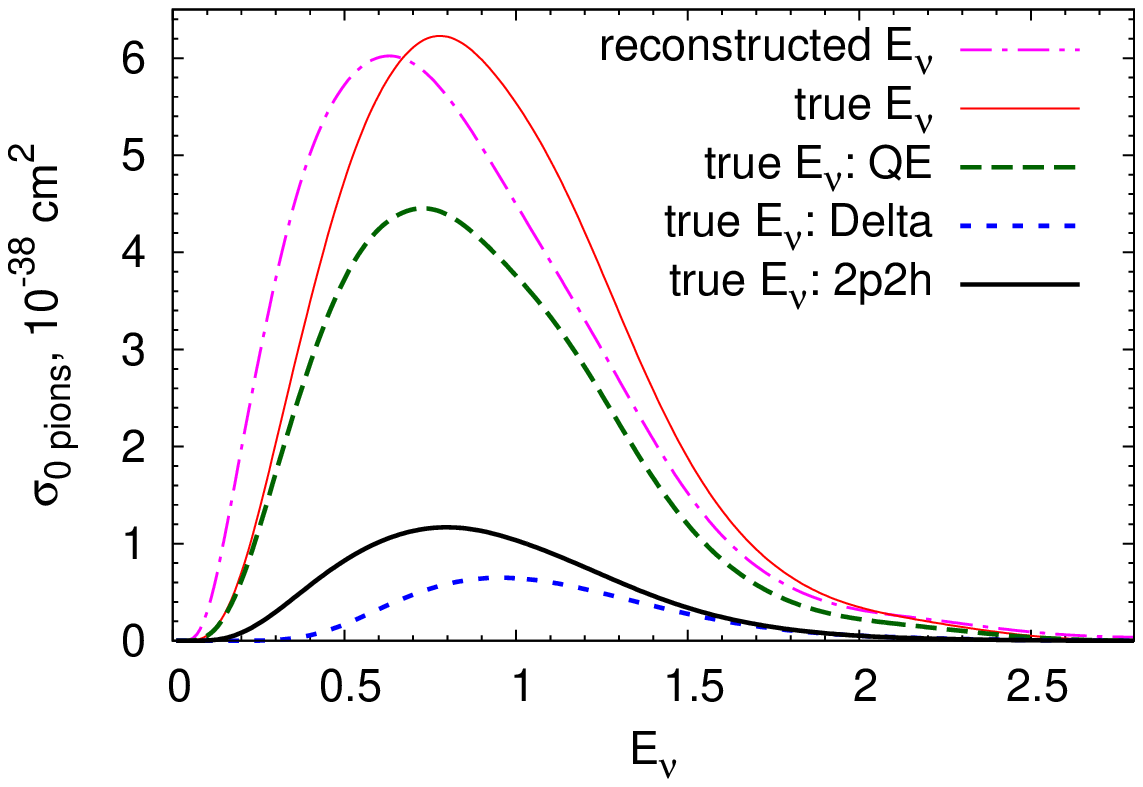}
\end{minipage}
\hfill
\begin{minipage}[c]{0.48\textwidth}
\includegraphics[scale=\plotscale]{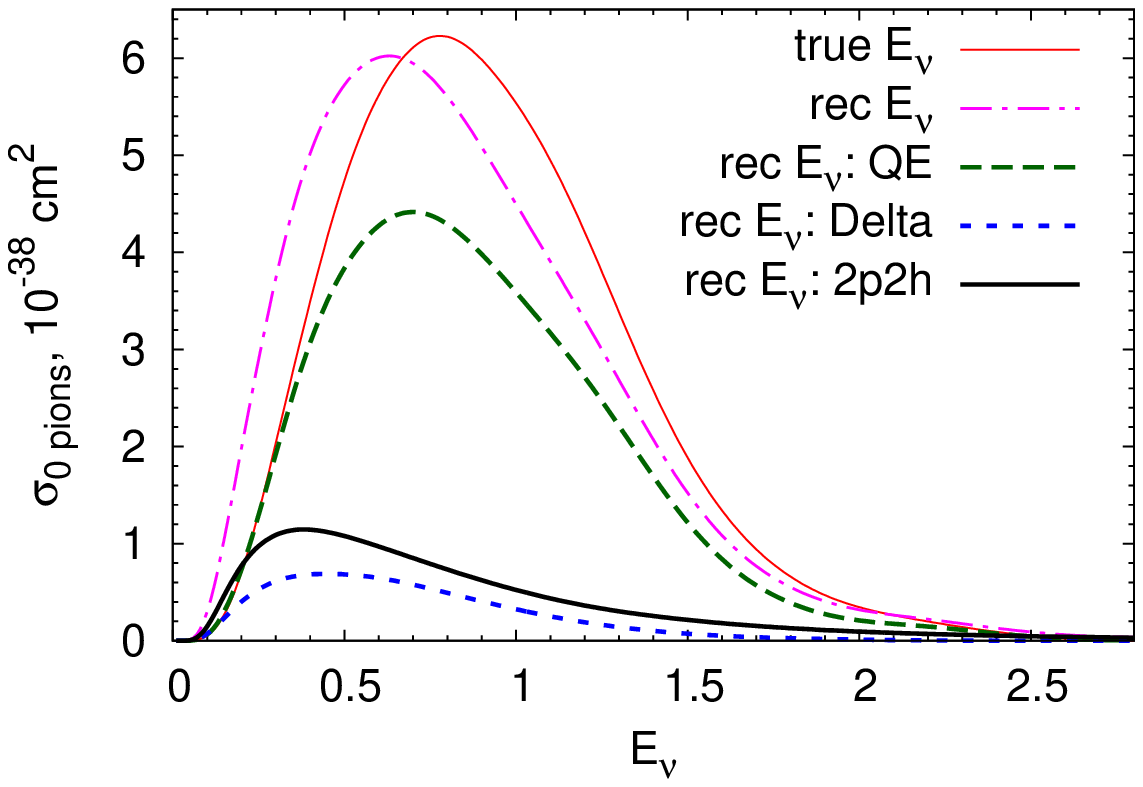}
\end{minipage}
\caption{(color online) Event distribution (i.e.\ the cross section multiplied by the neutrino flux) for QE-like processes as defined by the MiniBooNE experiment versus the real (left picture) and reconstructed (right picture) neutrino energy. The thin solid and dot-dashed curves repeat the distributions for true and reconstructed energy in both figures. Figure from \cite{Lalakulich:2012ac}.}
\label{fig:enu-reconstruction-MB-ME8}
\end{figure}

This is also seen in Fig.\ \ref{fig:enu-reconstruction-MB-ME8} which shows the event distribution (that is the cross section multiplied by the neutrino flux) for QE-like processes that would be observed by the MiniBooNE experiment. All events with zero pions in the final state are identified as QE-like ones. Since in the numerical simulation we know the true neutrino energy and the original interaction vertex, we can predict the origin of each event and its energy distribution for a given true neutrino energy. Only parts of these events are genuine QE events. The same QE-like events can originate from initial $\Delta$ production, higher resonance production (not shown in the figure), pion-background events (not shown in the figure) or 2p-2h processes. The event distribution versus the true energy is shown by the curve labeled 'true $E_\nu$'. Comparing it to the one versus the reconstructed energy (rec $E_\nu$) one sees that there is a systematic distortion in energy reconstruction: the peak of the event distribution is shifted now by about 100 MeV to \emph{lower} energies with this shift becoming smaller for the larger energies above the peak. Qualitatively this same behavior was already observed and discussed for the K2K and the MiniBooNE flux in \cite{Leitner:2010kp} as a consequence of pionic excitations. Here, now also the 2p-2h contribution is included in this analysis; it leads to the noticeable downward shift of the reconstructed energy curve on the low-energy side of the maximum where the pion production is still small.

\section{Summary}
In summary, understanding the reaction mechanism of neutrinos with nuclei is mandatory for a determination of neutrino oscillation parameters. Many of the relevant cross sections are experimentally not determined and have to be provided by theory. Here neutrino physics can only benefit from a close contact with photonuclear physics. This is true for the elementary cross sections, but it is also true for the all-important final state interactions. The latter are independent of the primary interaction process and can -- and must -- be checked with a multitude of other nuclear reactions. Among nuclear physicists working in heavy-ion physics and electron-, photon- or hadron-induced reactions on nuclei, there exists lots of experience and knowledge that still has to find its way into the neutrino community \cite{Mosel:2011ey}. In particular the all-important neutrino event-generators can only benefit from checks with the help of relevant photonuclear data which are readily available.

One important example for the influence of nuclear effects can be found in the determination of the neutrino energy which has to be known for the extraction of the neutrino mass- and mixing-parameters. Using quasielastic scattering as the main method to determine this energy requires a good identification of this reaction process. This involves the separation of 'clean' QE processes from resonance-excitations and many-particle interactions. This is illustrated very nicely in the results of the MiniBooNE experiment that have found widespread interest in the community over the last few years. The large excess of QE-like events observed by MiniBooNE over those calculated by commonly used event generators with the world-average value for the axial mass of around 1.0 GeV has led to a multitude of partly contradictory models that all aim to describe the same data starting from quite different initial 1p-1h or 2p-2h interaction mechanisms. Since broad energy-band neutrino experiments necessarily contain an averaging over quite different reaction mechanisms the inclusive data alone do not allow for an unambiguous experimental verification of any particular model. We have discussed that the energy reconstruction is affected by the presence of 2p-2h interactions mainly at the lower neutrino energies while the $\Delta$ excitations become more essential at the higher energies. 2p-2h interactions lead -- as the $\Delta$ excitations do -- to a downward shift of the reconstructed energy. Their contribution to the reconstructed energy distribution is fairly flat over a wide reconstructed energy range for the higher true energies.

In this talk we have given a short discussion of the relevant ingredients for an understanding of neutrino-nucleus data. Quasielastic scattering, pion production and many-particle interactions all have to be well controlled. The presently available inclusive data all rely for their understanding on event generators. A closer check can, however, probably come only from more exclusive data, like those on pion production and nucleon knock-out spectra. Nuclear physics can contribute a lot to understanding these data.

\section{Acknowledgements}
This talk and the article is based on material that was obtained in collaboration with Tina Leitner and Kai Gallmeister. We are grateful to them for a very productive, fruitful collaboration. We are also grateful to Alexei Larionov for many helpful detailed discussions.

\bibliographystyle{utphys}
\bibliography{nuclear,photonuclear}

\end{document}